\documentclass[a4paper,11pt]{article}
\usepackage{pos}
\usepackage{lineno}

\pdfoutput=1

\title{Measurement of inclusive jet production in Au+Au collisions at $\sqrt{s_\textrm{NN}}$ = 200 GeV by the STAR experiment}
\author*[a]{Robert Licenik (for the STAR Collaboration)}

\affiliation[a]{Nuclear Physics Institute, Czech Academy of Sciences,\\
  250 68, Rez, Czech Republic}
\emailAdd{licenik@ujf.cas.cz}

\abstract{The STAR Collaboration reports the measurement of inclusive jet production in central (0-10\%) and peripheral (60-80\%) Au+Au collisions at $\sqrt{s_\textrm{NN}}$~=~200~GeV, using both charged-particle and fully-reconstructed jets. Jet reconstruction is carried out using the anti-$k_\textrm{T}$ algorithm with resolution parameters $R$ = 0.2, 0.3 and 0.4. Yield suppression of charged-particle jets is observed for central Au+Au collisions relative to both peripheral Au+Au collisions and a vacuum baseline utilizing PYTHIA 6 simulations. The magnitude of the suppression is similar to that measured at the LHC and can be described by theoretical calculations. No evidence of significant medium-induced jet broadening is observed, based on comparison of jet spectra at varying $R$. The yield suppression, when expressed as the jet transverse momentum shift corresponding to energy loss, is consistent in magnitude with coincidence measurements at RHIC based on direct-photon and hadron triggers. There is an indication of larger energy loss observed at the LHC.}

\FullConference{%
  HardProbes2020\\
  1-6 June 2020\\
  Austin, Texas}
  
\begin{document}
\maketitle

The Quark-Gluon Plasma (QGP) created in high-energy heavy-ion collisions is opaque to jets (jet quenching) \cite{QGPreview}, a phenomenon that was first observed at RHIC via suppression of high transverse momentum ($p_\textrm{T}$) hadron yields and correlations \cite{PhysRevLett.91.072304}. Such measurements only provide limited insight into jet quenching mechanisms and dynamics, and more detailed measurements with reconstructed jets are required. While charged-particle and fully-reconstructed inclusive jet productions have been extensively studied in Pb+Pb collisions at the LHC (\cite{ALICEjets, ATLASjets, CMSjets}), these proceedings focus on inclusive jet production in heavy-ion collisions at RHIC. 
We discuss recently reported measurements of charged-particle jets in Au+Au collisions by the STAR Collaboration \cite{STARjets}, together with a new analysis to measure fully-reconstructed jets which is expected to have greater kinematic reach and improved systematic precision. The analysis uses data from the STAR detector \cite{STAR}, a large-acceptance system utilizing a solenoidal magnetic field, a Time Projection Chamber (TPC) \cite{TPC} for charged-particle tracking and momentum reconstruction, and the Barrel Electromagnetic Calorimeter (BEMC) \cite{BEMC}, which measures energy deposited by neutral particles and provides online triggers. STAR offers a full azimuthal coverage within pseudorapidity acceptance $|\eta| < 1$. The charged-jet analysis utilizes a dataset for Au+Au collisions at $\sqrt{s_\textrm{NN}}$~=~200~GeV with $L_\mathrm{int}~= 70~\upmu$b$^{-1}$, recorded in 2011 with a Minimum-Bias trigger. The fully-reconstructed jet analysis uses %equivalent of $L_{int}~= 5.2$ nb$^{-1}$ 
Au+Au collisions at $\sqrt{s_\textrm{NN}}$~=~200~GeV recorded in 2014 using a High-Tower trigger, which requires at least $\sim$ 4 GeV in one BEMC tower. %For details on the event selection see \cite{STARjets}.
% Events are accepted if their primary vertex was reconstructed within 30 cm from the nominal center of the STAR detector in the longitudinal direction ($V_z^{TPC} < 30 $ cm) and 2 cm in the perpendicular plane ($V_r < 2$ cm). In order to reduce pile-up, $|V_z^{TPC} - V_z^{VPD}| < 3$ cm, where $V_z^{VPD}$ is the position of the primary vertex along the beamline as determined by the VPD as opposed to the TPC, is also required. 
%Since there are no currently available results from $p+p$ collisions at $\sqrt{s}$~=~200~GeV with sufficient statistical precision, we use PYTHIA 6.428, tune Perugia 2012 (further tuned by STAR \cite{STARPYTHIA}) to simulate the vacuum reference for our measurements.
 
Details of the charged-particle jet analysis, based on charged-particle tracks measured in the TPC, are found in \cite{STARjets}. The fully-reconstructed jet analysis also utilizes BEMC clusters (3x3 adjacent towers), corrected for hadronic energy deposition. The cluster transverse energy is limited to $0.2 < E_\textrm{T} < 30.0$ GeV. %The jets are reconstructed using the anti-$k_\textrm{T}$ algorithm implemented in the FastJet package \cite{fastjet} with resolution parameter R = 0.2, 0.3 and 0.4. In order to assure that the whole jet is contained within the STAR detector acceptance, we require $|\eta_{jet}| < 1-$R. The jet area $A$ is also calculated by FastJet, using ghost particles. Jet reconstruction using the $k_\textrm{T}$ algorithm with R = 0.3 was carried out to determine the median background density $\rho = \textrm{median}\left\lbrace p_\textrm{T, jet}^\textrm{raw}/A_\textrm{jet}\right\rbrace$,
%where all jet candidates in the event, except for the (two) hardest jet candidate(s) in (peripheral) central collisions, were considered. The reconstructed jet momentum $p_\textrm{T, jet}^\textrm{raw}$ is then corrected for the underlying event contribution event-wise: $p_\textrm{T, jet}^\textrm{reco} = p_\textrm{T, jet}^\textrm{raw} - \rho \cdot A.$
%In order to remove large portion of purely combinatorial jets with very small area, jet candidates are required to satisfy jet area cut. The value of this cut is $A > 0.07/0.2/0.4$ for R = 0.2/0.3/0.4.
The combinatorial background in both analyses is removed by imposing a cut on the leading hadron transverse momentum $p_\textrm{T,lead}$. However, this cut also biases the fragmentation of the surviving jet population. This bias is measured by varying the $p_\textrm{T,lead}$ cut and results are presented for the unbiased region. 

\begin{figure}[ht!]
\begin{center}
\includegraphics[width=0.45\textwidth]{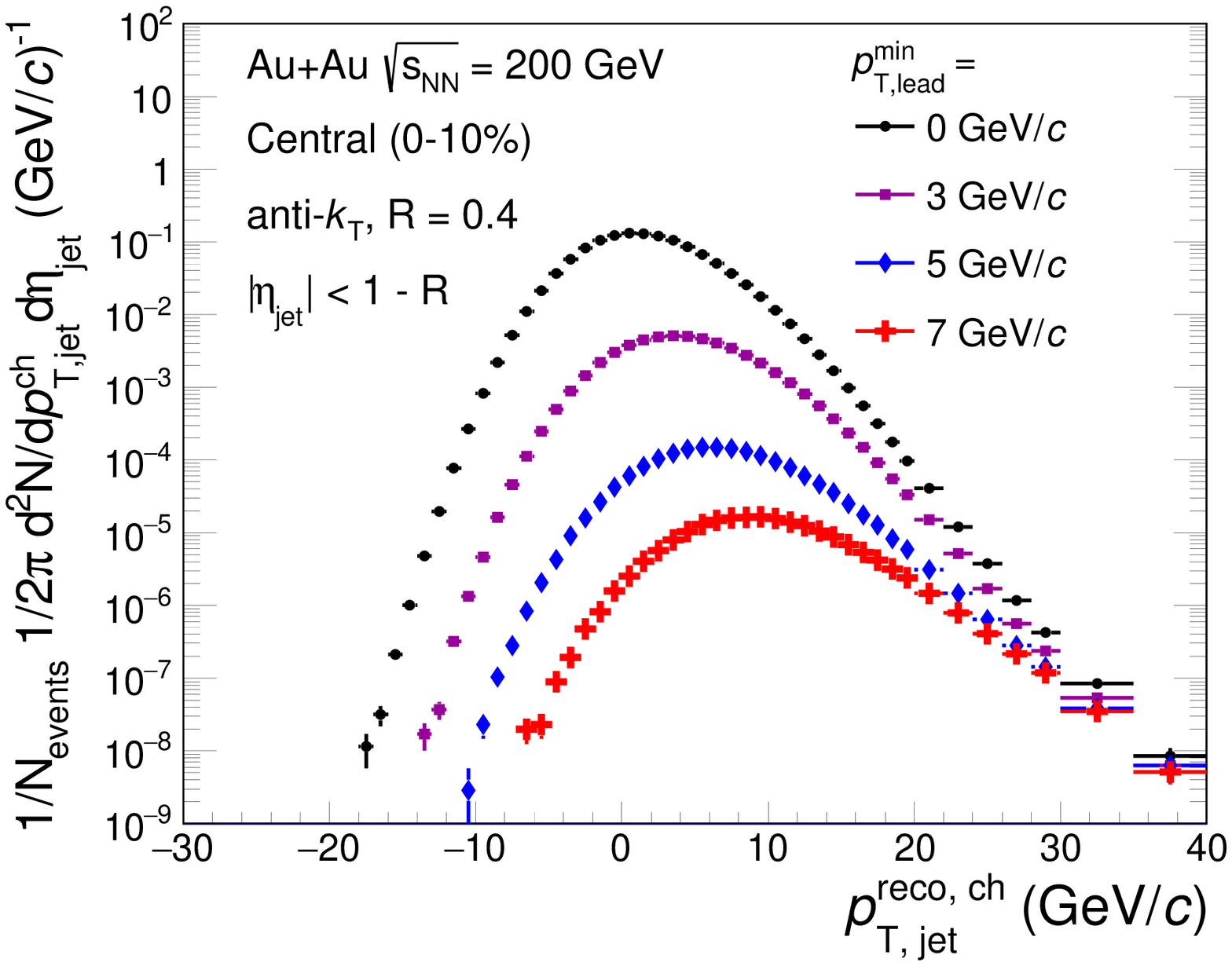}
\includegraphics[width=0.5\textwidth]{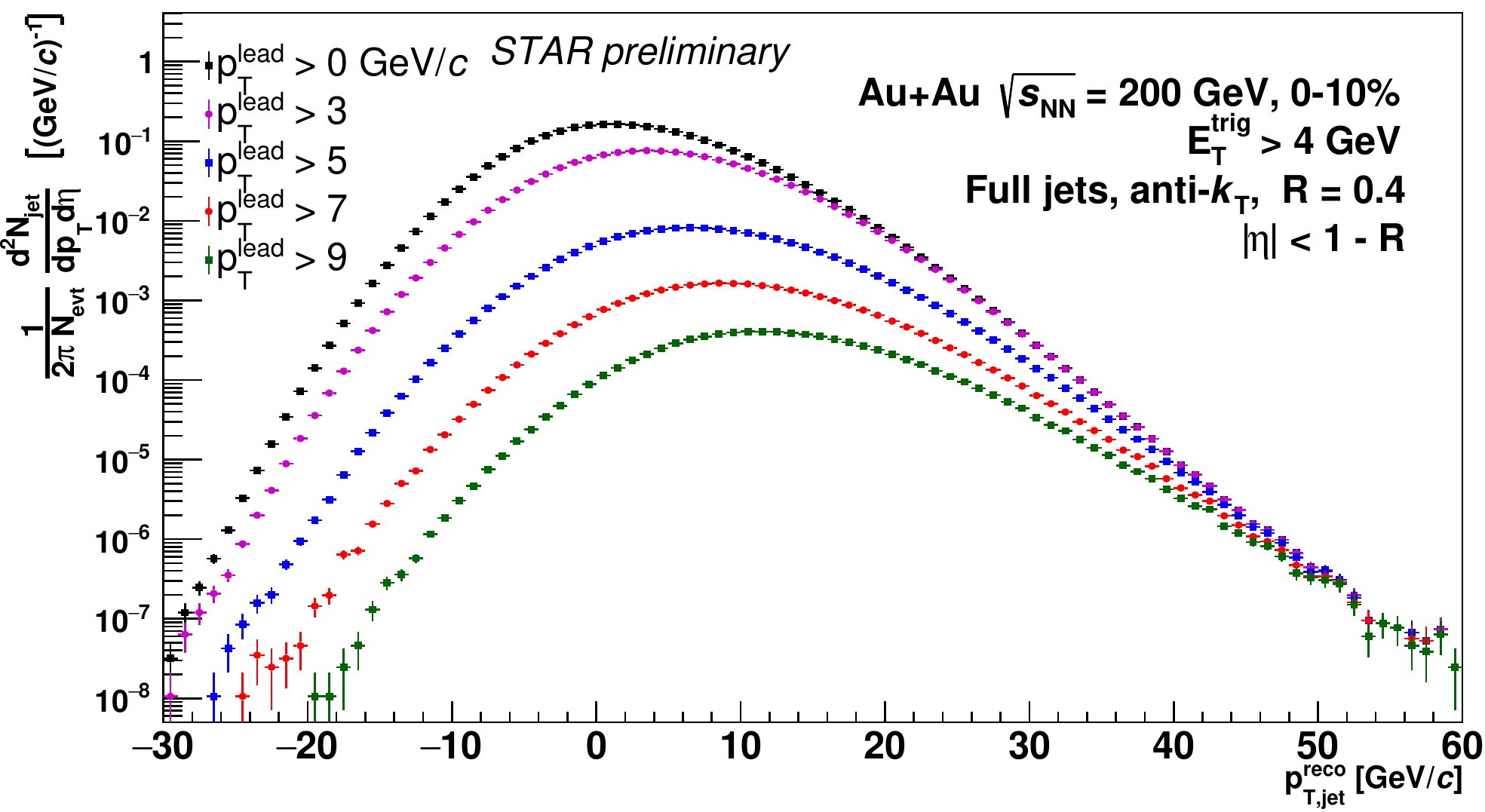}
\caption{Uncorrected distribution of charged-particle \cite{STARjets} (left) and fully-reconstructed (right) jets as a function of $p_\textrm{T,jet}^\textrm{reco}$ in 0-10\% Au+Au collisions at $\sqrt{s_\textrm{NN}}$~=~200~GeV. Different colors represent different values of the $p_\textrm{T,lead}$ cut.}
\label{fig: rawspectra}
\end{center}
\end{figure}

Figure \ref{fig: rawspectra} shows charged-particle (left) and fully-reconstructed (right) jet distributions as a function of $p_\textrm{T,jet}^\textrm{reco}$ ($= p_\textrm{T,jet}^\textrm{raw} - \rho \cdot A$, where $A$ is the jet area and $\rho$ is the median background energy density, calculated event-wise) for $R$ = 0.4 and various values of the $p_\textrm{T,lead}$ cut in central Au+Au collisions at $\sqrt{s_\textrm{NN}}$~=~200~GeV. It can be seen that the $p_\textrm{T,lead}$ cut significantly suppresses the combinatorial background, especially at low $p_\textrm{T,jet}^\textrm{reco}$. The distributions from the fully-reconstructed-jet analysis also indicate its extended kinematic reach, but corrected results are a work in progress. In the following we only show corrected results from the charged-particle jet analysis. Corrections are applied for the smearing effects of combinatorial background and instrumental effects using the SVD and Bayesian unfolding methods (details in \cite{STARjets}).

\begin{figure}[ht!]
\begin{center}
\includegraphics[width=\textwidth]{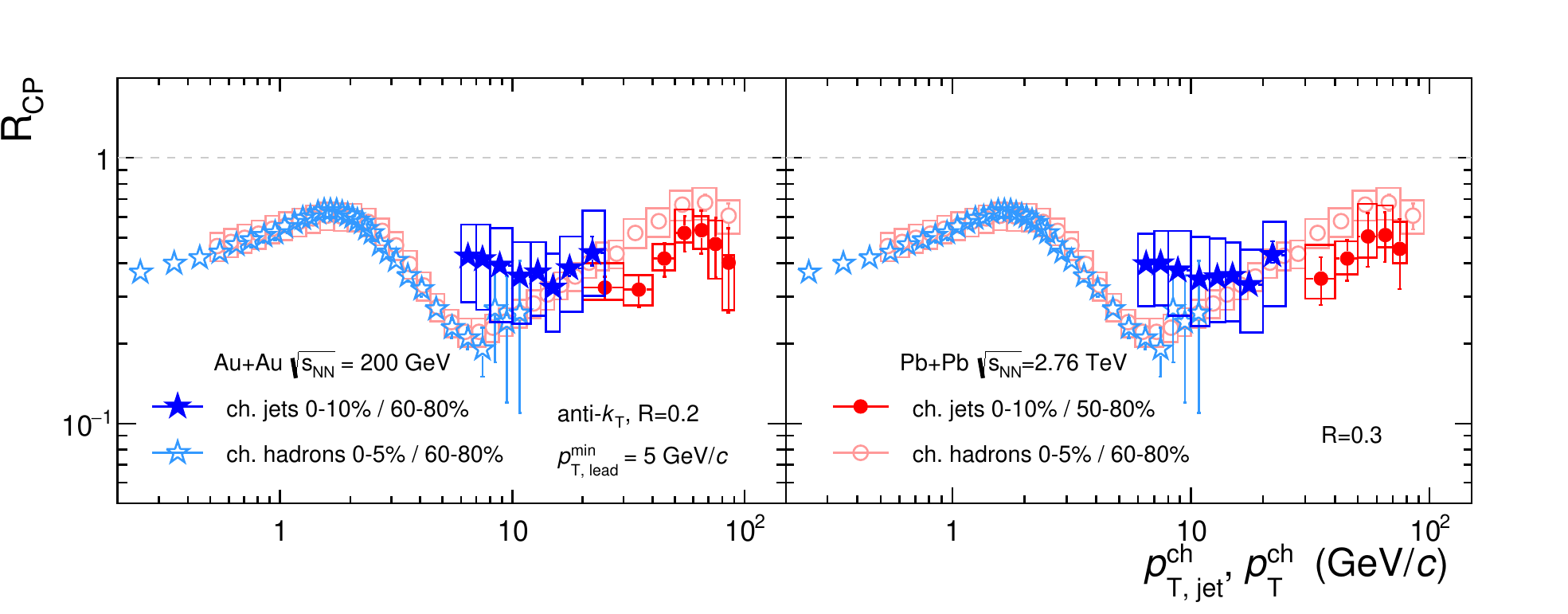}
\caption{$R_\mathrm{CP}$ of charged-particle jets reconstructed with $R$ = 0.2 and 0.3 and $p_\textrm{T,lead} > 5$ GeV/$c$ (solid stars) \cite{STARjets}. Also shown are similar suppression measurements with jets at the LHC \cite{ALICEjets} and inclusive charged hadrons at RHIC \cite{STARhadrons} and the LHC \cite{ATLAShadrons}.}
\label{fig: RCP}
\end{center}
\end{figure}

\begin{figure}[htp!] 
\begin{center}
\includegraphics[width=\textwidth]{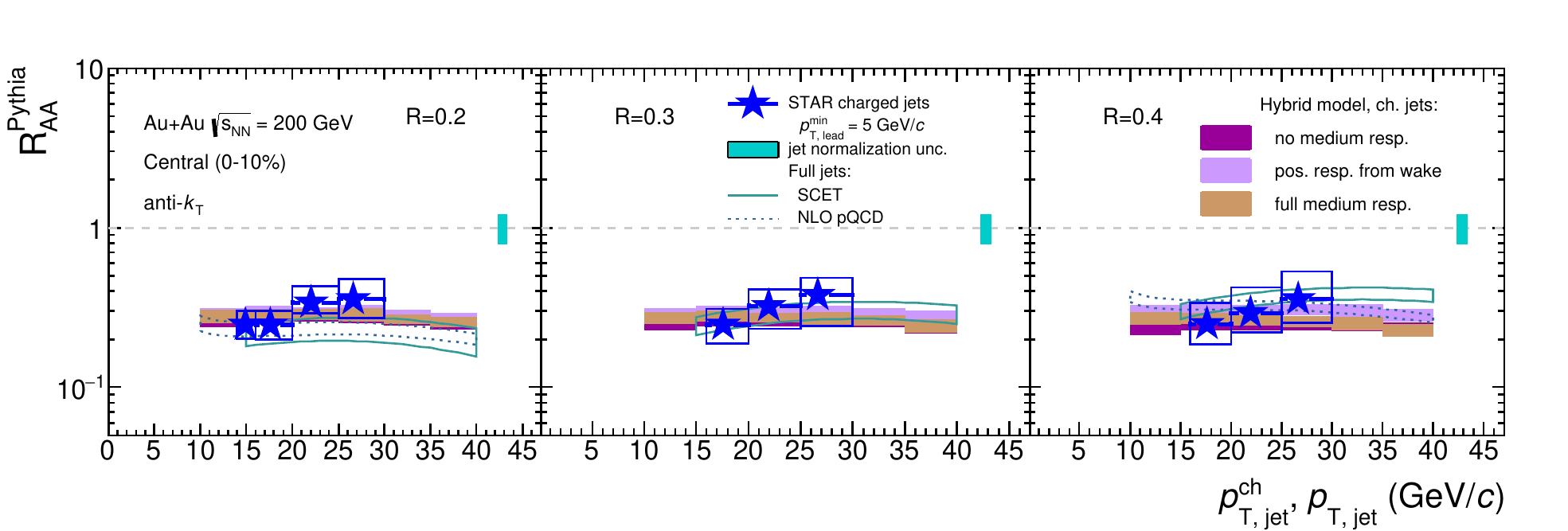}
\caption{$R^\mathrm{PYTHIA}_\mathrm{AA}$ as a function $p^\textrm{ch}_\textrm{T,jet}$ of charged-particle jets at STAR reconstructed with $R$ = 0.2, 0.3 and 0.4, and $p_\textrm{T,lead} > 5$ GeV/$c$  \cite{STARjets}. Bands represent theory predictions.}
\label{fig: RAA}
\end{center}
\end{figure}

Figure  \ref{fig: RCP} shows charged-particle jet $R_\mathrm{CP}$, the scaled ratio of yields in central to peripheral collisions, which exhibits a similar level of suppression to charged hadrons at RHIC \cite{STARhadrons} and LHC energies \cite{ATLAShadrons} and to charged-particle jets at the LHC at higher $p_\textrm{T,jet}$ \cite{ALICEjets}, with weak $p^\mathrm{ch}_\textrm{T,jet}$ dependence. Figure \ref{fig: RAA} shows charged-particle jet $R^\mathrm{PYTHIA}_\mathrm{AA}$, the yield suppression for central Au+Au collisions compared to $pp$ baseline calculated by PYTHIA 6 (Perugia 2012, further tuned by STAR \cite{STARPYTHIA}). Calculations based on jet quenching models \cite{VitevNLO, SCET1, SCET2, Hybrid}, shown in the various colored lines and shaded regions, are consistent with the measured value of $R^\mathrm{PYTHIA}_\mathrm{AA}$.

\begin{figure}[ht!]
\begin{center}
\includegraphics[width=0.8\textwidth]{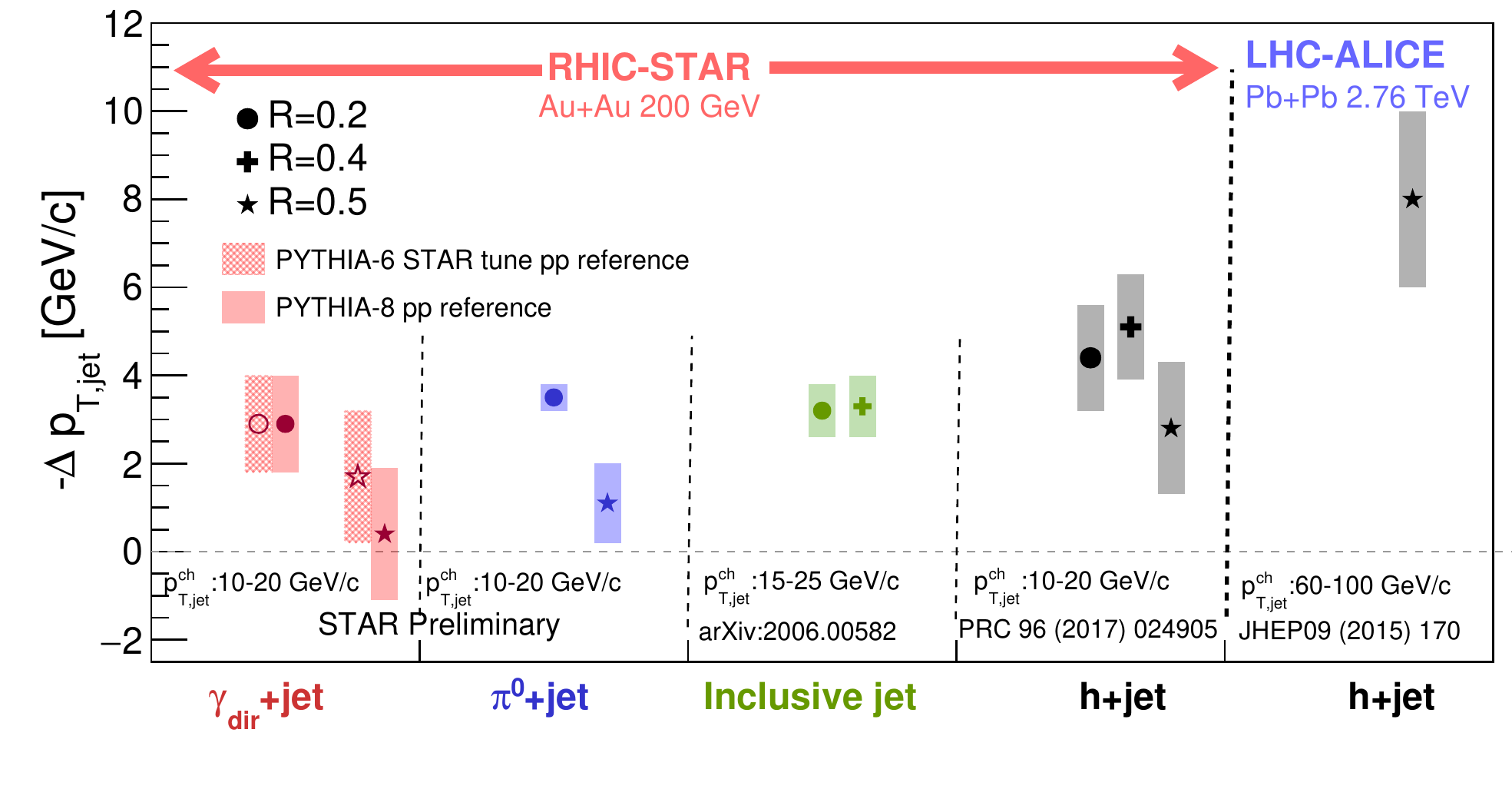}
\caption{Transverse momentum shift $-\Delta p_\textrm{T,jet}$ from this analysis (middle) compared to various semi-inclusive jet results at RHIC and LHC energies (references in figure).}
\label{fig: momentumshift}
\end{center}
\end{figure}

Figure \ref{fig: momentumshift} shows the transverse momentum shift $-\Delta p_\textrm{T,jet}$, corresponding to yield suppression \cite{STARhjet}, from neutral trigger+jet coincidence measurements at RHIC (red and blue points), inclusive jet measurement (green, this analysis) and charged hadron+jet coincidence measurements at RHIC and the LHC (black points). Results are consistent between channels at RHIC, and indicate smaller jet energy loss at RHIC than at the LHC (though the \textit{relative} shift appears larger at RHIC).

The ratio of inclusive jet cross-sections at different $R$ and fixed $p_\textrm{T,jet}$ measures the jet transverse energy profile. We do not observe any modification of transverse jet profile compared to $pp$ collision reference in peripheral or central collisions \cite{STARjets}. Dispersion of the model predictions is larger in this observable than in the $R^\mathrm{PYTHIA}_\mathrm{AA}$, which implies strong physical motivation to improve systematic uncertainties and study fully-reconstructed jets.

This work is supported by the project LTT18002 of the Ministry of Education, Youth and Sport of the Czech Republic.

\newpage
\bibliographystyle{JHEP}
%\bibliography{biblio.bib}{}

\begin{thebibliography}{10}

\bibitem{QGPreview}
W.~Busza, K.~Rajagopal and W.~van~der Schee, \emph{Heavy ion collisions: The
  big picture and the big questions},
  \href{https://doi.org/10.1146/annurev-nucl-101917-020852}{\emph{Annual Review
  of Nuclear and Particle Science} {\bfseries 68} (2018) 339–376}.

\bibitem{PhysRevLett.91.072304}
{\scshape STAR} collaboration, \emph{Evidence from $d+\mathrm{A}\mathrm{u}$
  measurements for final-state suppression of high-${p}_{T}$ hadrons in
  $\mathrm{A}\mathrm{u}+\mathrm{A}\mathrm{u}$ collisions at {RHIC}},
  \href{https://doi.org/10.1103/PhysRevLett.91.072304}{\emph{Phys. Rev. Lett.}
  {\bfseries 91} (2003) 072304}.

\bibitem{ALICEjets}
{\scshape ALICE} collaboration, \emph{{Measurement of charged jet suppression
  in {Pb-Pb} collisions at $ \sqrt{s_{\mathrm{NN}}} $ = 2.76 TeV}},
  \href{https://doi.org/10.1007/jhep03(2014)013}{\emph{Journal of High Energy
  Physics} {\bfseries 2014} (2014) }.

\bibitem{ATLASjets}
{\scshape ATLAS} collaboration, \emph{{Measurement of the nuclear modification
  factor for inclusive jets in Pb+Pb collisions at
  $\sqrt{{s}_{\mathit{\text{NN}}}}=5.02$ TeV with the ATLAS detector}},
  \href{https://doi.org/10.1016/j.physletb.2018.10.076}{\emph{Physics Letters
  B} {\bfseries 790} (2019) 108–128}.

\bibitem{CMSjets}
{\scshape CMS} collaboration, \emph{{Measurement of inclusive jet cross
  sections in $pp$ and PbPb collisions at
  $\sqrt{{s}_{\mathit{\text{NN}}}}=2.76$ TeV}},
  \href{https://doi.org/10.1103/PhysRevC.96.015202}{\emph{Phys. Rev. C}
  {\bfseries 96} (2017) 015202}.

\bibitem{STARjets}
{\scshape STAR} collaboration, J.~Adam, L.~Adamczyk, J.~R. Adams, J.~K. Adkins,
  G.~Agakishiev, M.~M. Aggarwal et~al., \emph{{Measurement of inclusive
  charged-particle jet production in {Au+Au} collisions at $\sqrt{s_{NN}}$=200
  GeV}},  2020.

\bibitem{STAR}
K.~Ackermann, N.~Adams, C.~Adler, Z.~Ahammed, S.~Ahmad, C.~Allgower et~al.,
  \emph{{STAR} detector overview},
  \href{https://doi.org/https://doi.org/10.1016/S0168-9002(02)01960-5}{\emph{Nuclear
  Instruments and Methods in Physics Research Section A: Accelerators,
  Spectrometers, Detectors and Associated Equipment} {\bfseries 499} (2003) 624
  }.

\bibitem{TPC}
M.~Anderson et~al., \emph{{{The STAR time projection chamber: A Unique tool for
  studying high multiplicity events at RHIC}}},
  \href{https://doi.org/10.1016/S0168-9002(02)01964-2}{\emph{Nucl. Instrum.
  Meth.} {\bfseries A499} (2003) 659}
  [\href{https://arxiv.org/abs/nucl-ex/0301015}{{\ttfamily nucl-ex/0301015}}].

\bibitem{BEMC}
M.~Beddo, E.~Bielick, T.~Fornek, V.~Guarino, D.~Hill, K.~Krueger et~al.,
  \emph{{The STAR Barrel Electromagnetic Calorimeter}},
  \href{https://doi.org/10.1016/S0168-9002(02)01970-8}{\emph{Nuclear
  Instruments and Methods in Physics Research Section A: Accelerators,
  Spectrometers, Detectors and Associated Equipment} {\bfseries 499} (2002)
  725}.

\bibitem{STARhadrons}
{\scshape STAR} collaboration, \emph{Transverse-momentum and collision-energy
  dependence of high-${p}_{T}$ hadron suppression in
  $\mathrm{A}\mathrm{u}+\mathrm{A}\mathrm{u}$ collisions at ultrarelativistic
  energies}, \href{https://doi.org/10.1103/PhysRevLett.91.172302}{\emph{Phys.
  Rev. Lett.} {\bfseries 91} (2003) 172302}.

\bibitem{ATLAShadrons}
{\scshape ATLAS} collaboration, \emph{{Measurement of charged-particle spectra
  in Pb+Pb collisions at $ \sqrt{s_{\mathrm{NN}}} $ = 2.76 TeV with the ATLAS
  detector at the LHC}},
  \href{https://doi.org/10.1007/jhep09(2015)050}{\emph{Journal of High Energy
  Physics} {\bfseries 2015} (2015) }.

\bibitem{STARPYTHIA}
{\scshape STAR} collaboration, \emph{Longitudinal double-spin asymmetry for
  inclusive jet and dijet production in $pp$ collisions at $\sqrt{s}=510\text{
  }\text{ }\mathrm{GeV}$},
  \href{https://doi.org/10.1103/PhysRevD.100.052005}{\emph{Phys. Rev. D}
  {\bfseries 100} (2019) 052005}.

\bibitem{VitevNLO}
I.~Vitev and B.-W. Zhang, \emph{Jet tomography of high-energy nucleus-nucleus
  collisions at next-to-leading order},
  \href{https://doi.org/10.1103/PhysRevLett.104.132001}{\emph{Phys. Rev. Lett.}
  {\bfseries 104} (2010) 132001}.

\bibitem{SCET1}
Y.-T. Chien, A.~Emerman, Z.-B. Kang, G.~Ovanesyan and I.~Vitev, \emph{Jet
  quenching from qcd evolution},
  \href{https://doi.org/10.1103/PhysRevD.93.074030}{\emph{Phys. Rev. D}
  {\bfseries 93} (2016) 074030}.

\bibitem{SCET2}
Y.-T. Chien and I.~Vitev, \emph{Towards the understanding of jet shapes and
  cross sections in heavy ion collisions using soft-collinear effective
  theory}, \href{https://doi.org/10.1007/jhep05(2016)023}{\emph{Journal of High
  Energy Physics} {\bfseries 2016} (2016) }.

\bibitem{Hybrid}
J.~Casalderrey-Solana, D.~C. Gulhan, J.~G. Milhano, D.~Pablos and K.~Rajagopal,
  \emph{Angular structure of jet quenching within a hybrid strong/weak coupling
  model}, \href{https://doi.org/10.1007/jhep03(2017)135}{\emph{Journal of High
  Energy Physics} {\bfseries 2017} (2017) }.

\bibitem{STARhjet}
{\scshape STAR} collaboration, \emph{{Measurements of jet quenching with
  semi-inclusive hadron+jet distributions in $\text{Au}+\text{Au}$ collisions
  at $\sqrt{{s}_{NN}}=200$ GeV}},
  \href{https://doi.org/10.1103/PhysRevC.96.024905}{\emph{Phys. Rev. C}
  {\bfseries 96} (2017) 024905}.

\end{thebibliography}

\providecommand{\href}[2]{#2}\begingroup\raggedright\endgroup

\end{document}